\documentclass{aa}
\pdfoutput=1

\usepackage[utf8]{inputenc}
\usepackage[varg]{txfonts}
\usepackage{graphicx}
\usepackage{empheq}
\usepackage{natbib}
\bibpunct{(}{)}{;}{a}{}{,} 

\newcommand{\Nh}{N$_{\text{H}}$}
\newcommand{\logNh}{$\log(\text{N}_\text{H})$}
\newcommand{\pex}{\texttt{pexrav}}
\newcommand{\myt}{\texttt{MYTORUS}}

\newcommand{\change}[1]{#1}
\newcommand{\changeb}[1]{#1}

\begin{document}

\title{Mildly obscured active galaxies and the cosmic X-ray background}

\author{V. Esposito\inst{1,2}
        \and
        R. Walter\inst{1,2}
       }

\institute{ISDC Data Centre for Astrophysics, University of Geneva, ch. d'Ecogia 16, 1290 Versoix, Switzerland\\
          \email{Valentino.Esposito@unige.ch}
          \and
          Geneva Observatory, University of Geneva, ch. des Maillettes 51, 1290 Versoix, Switzerland
          }

\date{Received ...; accepted ...}

\abstract{The diffuse cosmic X-ray background (CXB) is the sum of the emission of discrete sources, mostly massive black-holes accreting matter in active galactic nuclei (AGN). The CXB spectrum differs from the integration of the spectra of individual sources, calling for a large population, undetected so far, of strongly obscured Compton thick AGN. Such objects are predicted by unified models, which attribute most of the AGN diversity to their inclination on the line of sight, and play an important role for the understanding of the growth of black holes in the early Universe.}
{The fraction of obscured AGN at low redshift can be derived from the observed CXB spectrum assuming AGN spectral templates and luminosity functions.}
{We show that high signal-to-noise average hard X-ray spectra, derived from more than a billion seconds of effective exposure time with the Swift/BAT instrument, imply that mildly obscured Compton thin AGN feature a strong reflection and contribute massively to the CXB.}
{A population of Compton thick AGN larger than that effectively detected is not required, as no more than 6\% of the CXB flux can be attributed to them. The stronger reflection observed in mildly obscured AGN suggests that the covering fraction of the gas and dust surrounding their central engines is a key factor in shaping their appearance. These mildly obscured AGN are easier to study at high redshift than Compton thick sources.}
{}



\maketitle

\section{Introduction\label{sec:introduction}}

The cosmic X-ray background (CXB) is a broad band diffuse emission discovered since the early era of X-ray astronomy \citep{1962PhRvL...9..439G}. It is known today that in the X-ray domain the CXB is the integrated emission of many faint point-like extragalactic sources, most of them being Seyfert galaxies or quasars.

At energies $< 10$ keV the CXB emission has been almost completely resolved into its constituting point-like sources \citep{2005MNRAS.357.1281W} thanks to the focusing capabilities of soft X-rays instruments like XMM-Newton and Chandra. The sum of the fluxes of the sources detected in deep X-ray surveys performed by XMM-Newton in the Lockman Hole \citep{2001A&A...365L..45H} and by Chandra \citep{2002ApJS..139..369G,2003AJ....126..539A} almost reaches the level of the CXB flux, confirming that the CXB is the combination of the emission of faint Active Galactic Nuclei (AGN) with different degrees of obscuration.

The sensitivity of current instruments is not enough to resolve the CXB at hard X-rays where lies the bulk of its emission. The extrapolated flux of the AGN population resolved at lower energies is not able to explain the CXB flux at $E \sim 30$ keV. The strength of X-ray reflection in AGN spectra is a possibile solution to this discrepancy. Moreover, the reflection component is the only relevant one in heavily obscured, Compton thick AGN (CTK), i.e. sources where the density of the obscuring material is high enough for Compton scattering to dominate (\Nh\ $> 10^{24}~\text{cm}^{-2}$).

The CXB flux and spectrum have been measured by ASCA/SIS \citep{1995PASJ...47L...5G}, ROSAT \citep{1998A&A...334L..13M}, RXTE/PCA \citep{2003A&A...411..329R}, XMM-Newton \citep{2004A&A...419..837D}, Chandra \citep{2006ApJ...645...95H} and Swift/XRT \citep{2009A&A...493..501M} at soft X-rays, and by HEAO1 \citep{1980ApJ...235....4M,1999ApJ...520..124G,2005A&A...444..381R} and more recently by \mbox{Beppo-SAX} \citep{2007ApJ...666...86F}, INTEGRAL \citep{2007A&A...467..529C,2010A&A...512A..49T} and Swift/BAT \citep{2008ApJ...689..666A} at hard X-rays. The measurements are in agreement at a level of $\sim 10 - 15\%$ throughout the full energy range of the CXB \citep{2005A&A...444..381R,2007ApJ...666...86F,2008ApJ...689..666A,2009A&A...493..501M}.

The fraction of AGN which need to be CTK to explain the CXB flux is still debated: e.g. \citet{2007A&A...463...79G} suggested that 50\% of the obscured AGN have to be CTK, a similar value is proposed also by \citet{2014ApJ...786..104U} while \citet{2009ApJ...696..110T} proposed a smaller fraction, around 15\%.

The method to perform the CXB synthesis has been developed in the seminal works of \citet{1989A&A...224L..21S,1995A&A...296....1C} and improved in following works. To synthesize the CXB spectrum three main ``ingredients'' must be known: an accurate description of the broad band spectra of the various AGN classes, the luminosity function which gives the number density of AGN per comoving volume as a function of luminosity and redshift, and the so called \Nh\ distribution, i.e. the distribution of AGN as a function of absorbing column density (\Nh).

AGN spectra are provided as spectral templates for the various AGN classes: previous works set the parameters of their spectral templates to values representative of observations \citep[e.g.][]{2007A&A...463...79G} or of models \citep[e.g.][]{2014ApJ...786..104U}. Special care should be used for the CTK template: to model the effect of Compton scattering, especially the scattered component, assumptions on the geometry of the scattering material must be taken, and the spectrum is usually modelled through Monte Carlo simulations assuming a torus geometry for the absorbing material according to the AGN Unified model, e.g. \citet{2011MNRAS.413.1206B,2014MNRAS.443.1999B} or the \myt\ model\footnote{http://mytorus.com} proposed by \citet{2012MNRAS.423.3360Y}.

The AGN X-ray luminosity function (XLF) is derived from deep surveys \citep[e.g.][]{2003ApJ...598..886U,2005ApJ...635..864L,2005A&A...441..417H,2010MNRAS.401.2531A,2014ApJ...786..104U,2015ApJ...804..104M,2015ApJ...802...89B,2015MNRAS.451.1892A,2015arXiv151205563R} and hence is available only in the soft X-ray range. The \Nh\ distribution can be derived from data \citep[e.g.][but it is biased against the detection of highly absorbed sources]{2007A&A...463...79G} or from models \citep[e.g.][and references therein]{2009ApJ...696..110T}.

As there are many parameters involved in the synthesis, there is a certain level of degeneration in the process, especially in the CTK determination, as shown by \citet{2009ApJ...696..110T} and \citet{2012A&A...546A..98A}. Both authors show that the CTK density in the Universe can not be constrained by the CXB alone, the main reason being the lack of a robust CTK spectral template.

In this work we use parameters derived from average spectra of AGN classes measured by Swift/BAT (14 -- 195 keV) to build spectral templates representative of real spectra, resolving the uncertainty in the synthesis process due to the choice of the templates. The shape of the templates strongly affect the synthetized CXB spectrum, specially at hard X-rays where lies the bulk of the CXB emission and where the contribution of CTK sources is relevant.

The AGN sample and the BAT stacking method are described in detail in Sections \ref{sec:sample} and \ref{sec:stacking}. In Sections \ref{sec:templates} and \ref{sec:xlf} we describe the ingredients used in the synthesis, i.e. the spectral templates, the XLFs and the \Nh\ distributions. In Section \ref{sec:results} we present the results and in particular the maximum fraction of CTK allowed by the data and we discuss them in Section \ref{sec:discussion}.

Throughout this paper we assume $H_0 = 70 \, \text{km} \, \text{s}^{-1} \, \text{Mpc}^{-1}$, $\Omega_M = 0.3$ and $\Omega_\Lambda = 0.7$.

\section{The Sample\label{sec:sample}}
We use the final sample of 165 sources of \citet{2011A&A...532A.102R} (R11) to build the stacked spectra. All these sources are Seyfert Galaxies detected by INTEGRAL IBIS/ISGRI for which redshift and \Nh\ measurements are available: the redshift spans from 0.001 to 0.162, the \Nh\ from $4~10^{19}$ to $2.1~10^{24}$ cm$^{-2}$.

All these sources have also been detected by Swift/BAT \citep{2010ApJS..186..378T,2013ApJS..209...14K}. As the BAT field of view is $\sim 10$ times larger than that of ISGRI, stacking BAT data allows to achieve unprecedent statistics.

R11 analysed stacked ISGRI spectra of Seyfert Galaxies, introducing a classification based on absorption: Seyfert 2 galaxies are divided into Lightly Obscured sources (LOB), Mildly Obscured sources (MOB) and CTK sources.

We notice that the \Nh\ distributions of these 4 samples are overlapping: there are 7 sources classified as Seyfert 1 with \Nh\ $> 10^{22}~\text{cm}^{-2}$, 2 Seyfert 2 galaxies with \Nh\ $< 10^{21}~\text{cm}^{-2}$, and that 29 sources, classified as Seyfert 1.5, were not included in these samples. Hence we defined an alternate set of samples based exclusively on the \Nh: Unabsorbed (\Nh~$<~10^{21}~\text{cm}^{-2}$), LOB1 ($10^{21} < \text{N}_\text{H} < 10^{22}~\text{cm}^{-2}$), LOB2 ($10^{22} < \text{N}_\text{H} < 10^{23}~\text{cm}^{-2}$), MOB ($10^{23} < \text{N}_\text{H} < 10^{24}~\text{cm}^{-2}$) and CTK (\Nh~$>~10^{24}~\text{cm}^{-2}$).

The total number of sources in each set is reported in the first column of Table \ref{tab:time}. These two sets of samples were used to produce stacked images to derive two different sets of spectral templates for the CXB synthesis.

\section{Stacking of BAT data\label{sec:stacking}}

\begin{table}
\caption{\change{Effective exposure and average count rate derived from the mosaic images for each set of stacked images. For each sample we reported the number of stacked sources, the effective exposure, the significance and the count rate.}}
\label{tab:time}
\begin{tabular}{lcccc}
\hline\noalign{\smallskip}
Sample & Sources & Eff. exp. & Signif. & Rate \\ 
       & & $10^9$ s  &  & ct s$^{-1}$ pix$^{-1}$ \\ 
\hline\noalign{\smallskip}
\multicolumn{5}{c}{Samples based on R11 definition} \\
\hline\noalign{\smallskip}
Seyfert 1     & 44 & 0.627 & 93 & 7.82 $10^{-6}$ \\ 
LOB Seyfert 2 & 34 & 0.478 & 103 & 9.43 $10^{-6}$ \\ 
MOB Seyfert 2 & 27 & 0.384 & 87 & 9.12 $10^{-6}$ \\ 
CTK           & 10 & 0.140 & 57 & 10.31 $10^{-6}$ \\ 
\hline\noalign{\smallskip}
\multicolumn{5}{c}{Samples based on \Nh} \\
\hline\noalign{\smallskip} 
Unabsorbed & 35 & 0.466 & 111 & 8.30 $10^{-6}$ \\
LOB1       & 26 & 0.388 & 85 & 7.40 $10^{-6}$ \\
LOB2       & 35 & 0.482 & 100 & 6.73 $10^{-6}$\\
MOB        & 29 & 0.392 & 91 & 7.13 $10^{-6}$ \\
CTK        & 10 & 0.140 & 57 & 10.31 $10^{-6}$ \\ 
\noalign{\smallskip}\hline
\end{tabular}
\end{table}

\change{The main scientific goal of the Burst Alert Telescope (BAT) on board Swift satellite \citep{2004ApJ...611.1005G} is the detection of hard X-ray transients, especially Gamma-Ray Bursts. Thanks to its large field of view (1.4 sr partially-coded) it is also an ideal survey instrument, observing the full sky for very long integrated exposure time.}

\changeb{The Swift/BAT detector consists of $2^{15}$ pixels of CdZnTe \citep{2005SSRv..120..143B} recording X-ray photons arriving from the sky through a random (50\% open) coded mask, made of $\sim 54000$ obscuring lead tiles supported by a honeycomb panel. Each source in the field of view projects a shadow of the mask on the detector plane. The on-board electronic accumulates 80-bin spectra from every pixels during a fixed integration period (typically 5 minutes) and send these histograms to ground, together with other data streams. The signal from each source and from the background (dominated by the CXB) can be reconstructed through an image deconvolution. Most sources can only be detected in mosaic of sky images obtained from many ($\sim 1000$ sec) spacecraft pointings.}

\change{The Swift/BAT reduction pipeline for the all-sky survey is described in \citet{2008ApJ...681..113T,2010ApJS..186..378T}, \citet{2013ApJS..207...19B} and \citet{2013ApJS..209...14K}. Our pipeline follows this closely and is based on the BAT analysis software HEASOFT v 6.13.}

\change{A first analysis was performed to derive background detector images. We created sky images (task \texttt{batsurvey}) in the 8 standard energy bands (in keV: 14 - 20, 20 - 24, 24 - 35, 35 - 50, 50 - 75, 75 - 100, 100 - 150, 150 - 195) using an input catalogue of 86 bright sources that have the potential to be detected in single pointings. The detector images were then cleaned by removing the contribution of all detected sources (task \texttt{batclean}) and averaged to obtain one background image per day. The variability of the background detector images was then smoothed pixel-by-pixel fitting the daily background values with different function (spline, polynomial). A polynomial model with an order equal to the number of months in the data set adequately represents the background variations. A similar background smoothing function was used by the BAT team.}

\change{The BAT image analysis was then run again using these smoothed averaged background maps. The new sky images were then stored in an all-sky pixel database by properly projecting the data \changeb{on a fixed grid of sky pixel}, preserving fluxes (the angular size of the BAT pixels varies in the field of view). This database can then be used to build local images and spectra or lightcurves for any sky position.}

\change{The result of our processing was compared to the standard results presented by the Swift team\footnote{http://swift.gsfc.nasa.gov/results/bs70mon/} for individual sources and a very good agreement was found.}\\

To extract the average spectra of the different samples of Seyfert galaxies, we followed the procedure adopted by \citet{2009A&A...497...97W} and R11. We created $500\times500$-pixels mosaic images modifying the coordinate system of each individual image, setting the coordinates of each source of the sample to an arbitrary fixed position $(\alpha=0,  \delta=0)$. The geometry of the image was also modified to obtain a uniform point spread function at the center of the mosaic whatever is the position of the source in the field of view. These mosaic images, built independently for each energy band, provide a stack of all the sources of a given sample.

To minimize the systematics in the mosaic images, we excluded the noisiest images. A total of 201130 sky images were finally included in the processing, \changeb{covering 8 years from January 2005 to December 2012.} \change{Each sky image was in fact included many times, once for each source of the sample present in its field of view.} The mosaic images were built with a tangential projection using a factor of two oversampling when compared to the individual input sky images. This results in a pixel size of 7.5954 arcmin at the center of the mosaics. The photometric integrity and accurate astrometry were obtained by calculating the intersection between input and output pixels, and weighting the count rates according to the overlapping area.

The average signal, extracted from the mosaic for each individual sample, the exposures and the number of sources used are reported in Table \ref{tab:time}. The detection significances \change{of the stacked sources in the mosaics} \changeb{are calculated as the number of counts divided by the square root of the variance and} range between $111\sigma$ and $57\sigma$. The effective exposure obtained at the center of the mosaics are between 0.627 Gs and 0.140 Gs.

\begin{figure}
\centering
\includegraphics[width=\hsize]{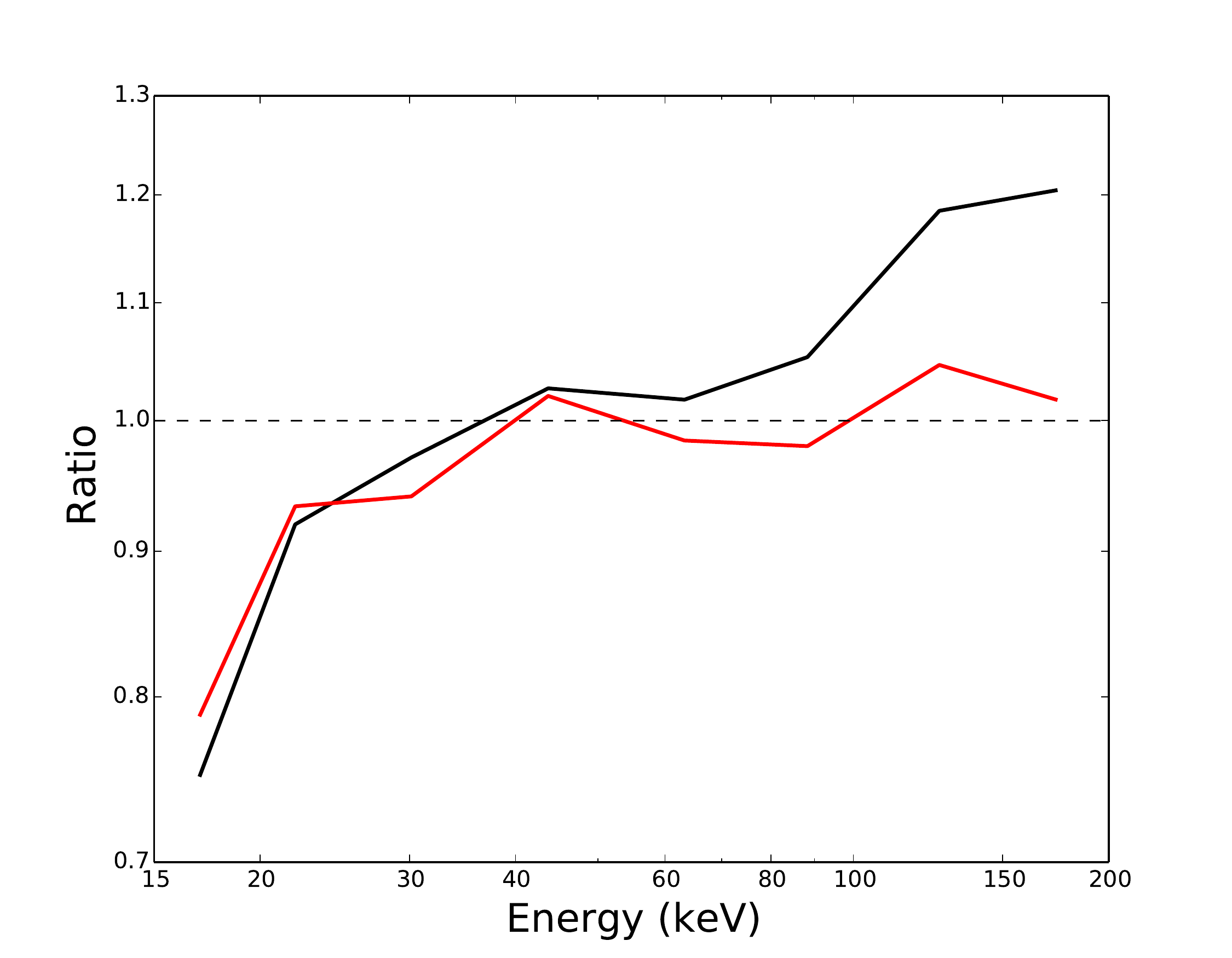}
\caption{\changeb{Ratio between the averaged Crab spectrum and a powerlaw \citep[red: our work; black: from][]{2013ApJS..207...19B}, using an on-axis response.}}
\label{fig:matrixcorr}
\end{figure}

The eight bin stacked spectrum for each AGN sample was extracted from the mosaic images using \texttt{mosaic\_spec}\footnote{from the OSA package, http://isdc.unige.ch/integral/analysis}. \changeb{The standard BAT response matrix depends on the off-axis angle and cannot be used when data from many off-axis angles are averaged. To produce a suitable matrix we followed the same procedure adopted by the BAT team for the survey work \citep{2013ApJS..207...19B}, which consists in tuning the on-axis response to obtain a correct spectrum for the Crab nebula. The tuning accounts for the fact that a smaller fraction of the low energy photons are detected on average over the field of view when compared with the on-axis expectation, largely because of absorption in the honeycomb mask supporting structure. The correction therefore depends on the average spectrum of the Crab nebula. Figure \ref{fig:matrixcorr} shows the ratio between the observed spectrum \cite[obtained by us and by ][]{2013ApJS..207...19B} and the powerlaw (F$_\nu \sim \nu^{-1.15})$ model representing the Crab nebula, obtained assuming the on-axis response. The effect of the absorption at low energy is clearly observed. Up to 80 keV the observed spectra are consistent within a few \%. At higher energies \cite{2013ApJS..207...19B} required an additional correction that we are not observing. The red curve in figure \ref{fig:matrixcorr} is very similar to the NOMEX absorption model used for the supporting structure of the IBIS mask on board INTEGRAL \citep{2006privcommLub}. We finally used the response matrix provided by \cite{2013ApJS..207...19B}\footnote{http://swift.gsfc.nasa.gov/results/bs70mon/}, modified to obtain a good powerlaw adjustment to the spectrum of the Crab nebula, in particular above 80 keV. A systematic uncertainty of 2\% has been added in all spectral model adjustments to account for the uncertainty related to the off-axis effects at low energies. We also built a second response matrix assuming a broken powerlaw for the Crab spectrum \citep{2008int..workE.144J} and found that the resulting spectral parameters (Tab. \ref{tab:spectralparameters}) were identical within the uncertainties.}

\section{Stacked spectra and spectral templates\label{sec:templates}}

The stacking of the Swift/BAT data described in Sec. \ref{sec:stacking} allows us to obtain the average spectra of the Seyfert galaxy classes in the BAT energy bandpass (14 -- 195 keV) characterized by different absorption levels. The parameters derived from these spectra are then used as spectral templates in the CXB synthesis.

The stacked spectra of the samples built for \logNh\ $< 24$ have been fitted with the \pex\ model modified for photoelectric absorption. The \pex\ model consists of an exponentially cut off powerlaw plus reflection from an infinite slab of neutral material, representing an accretion disk \citep{1995MNRAS.273..837M}. The absorption has been fixed to the central value (in logarithmic space) of the included sources.

\begin{figure}
\centering
\includegraphics[width=\hsize]{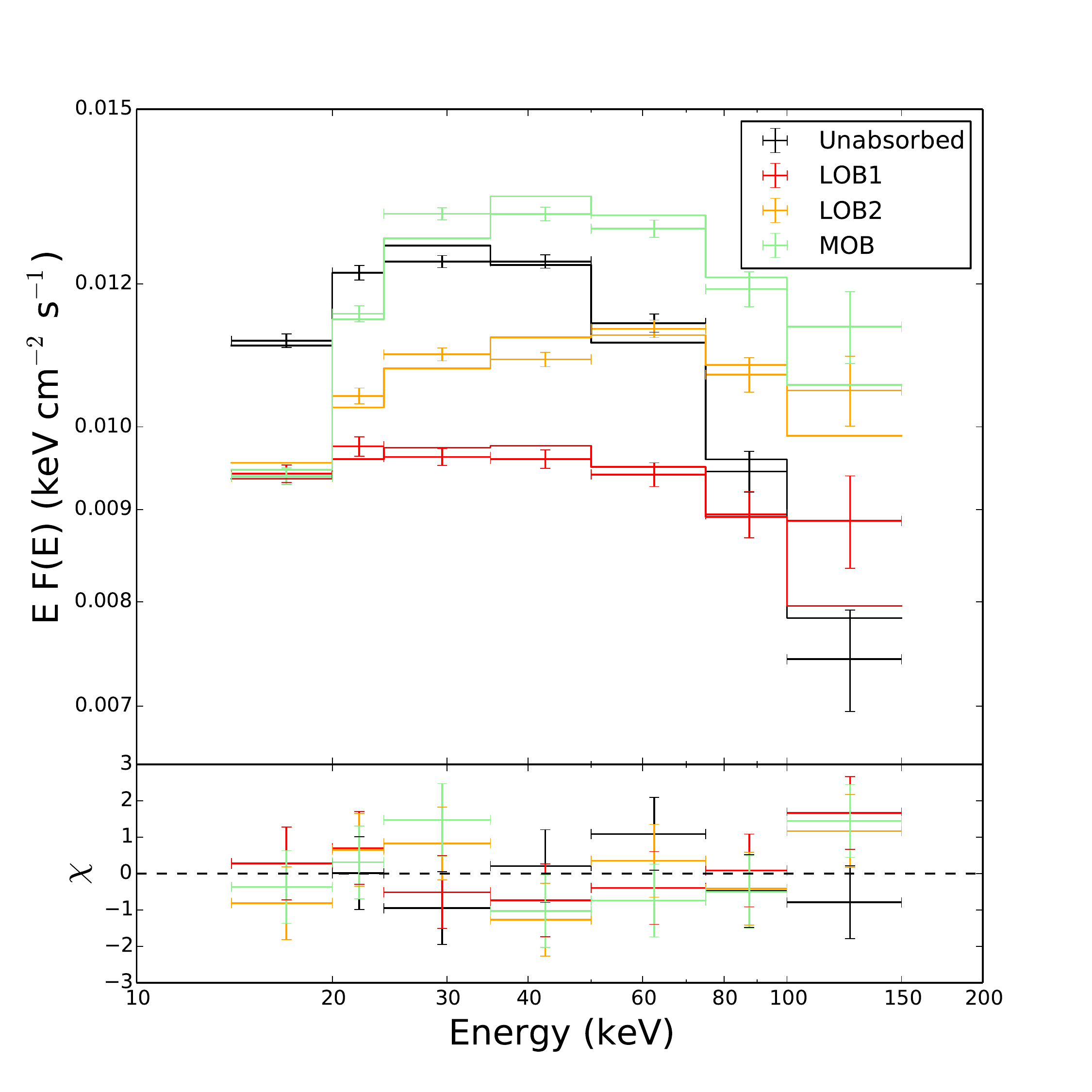}
\caption{Swift/BAT stacked spectra of the samples based on \Nh\ and fitted with \texttt{phabs * pexrav} model.}
\label{fig:spectra}
\end{figure}

\change{The spectral fit has been performed with \texttt{XSPEC} (version 12.8.2) using the $\chi^2$ minimization method. As described in Section \ref{sec:stacking} the spectra are binned in 8 energy bins, and the stacked sources are not detected in the highest one, so it is ignored in the spectral fit. The spectra are shown in Figure \ref{fig:spectra} with the best fit model and residuals.}

The impressive statistics \changeb{(the total number of photons in the spectra ranges between $\sim 12000$ and $\sim 34000$)} obtained with the stacking method allows us to measure with good accuracy the relevant spectral parameters, i.e. the spectral index $\Gamma$ and the covering factor of the reflecting material R. Table \ref{tab:spectralparameters} reports these parameters for each spectrum. The cutoff energy $E_c$ lies beyond the upper threshold of the BAT energy band and can not be constrained by the fit, so we fixed it at $E_c = 200$ keV, as it is unlikely to have higher values \citep{2014ApJ...782L..25M}. \change{We fixed also the inclination angle $\theta$ at the value of $\cos\theta = 0.45$} and solar abundancies were assumed.

\begin{table}
\caption{\change{Spectral parameters (spectral index $\Gamma$ and reflection R) derived from the fits of the stacked spectra and used for the spectral templates of the CXB synthesis. 1 $\sigma$ errors derived from the fits are reported in the last two columns. The absorption is always a fixed parameter.}}
\label{tab:spectralparameters}
\begin{tabular}{lccccc}
\hline\noalign{\smallskip}
Sample & \logNh & $\Gamma$ & R & $\Delta \Gamma$ & $\Delta$R \\
\hline\noalign{\smallskip}
\multicolumn{6}{c}{Stacked spectra based on R11 definition} \\
\hline\noalign{\smallskip} 
Seyfert 1     & 0 & 1.86 & 0.17 & 0.02 & 0.14 \\ 
\noalign{\smallskip}
LOB Seyfert 2 & 22 & 1.68 & 0.0 & 0.02 & 0.04 \\ 
\noalign{\smallskip}
MOB Seyfert 2 & 23.5 & 1.74 & 0.88 & 0.02 & 0.12 \\ 
\hline\noalign{\smallskip} 
\multicolumn{6}{c}{Stacked spectra based on \Nh} \\
\hline\noalign{\smallskip}
Unabsorbed & 0 & 1.89 & 0.77 & 0.02 & 0.2 \\
\noalign{\smallskip}
LOB1       & 21.5 & 1.82 & 0 & 0.02 & 0.04 \\
\noalign{\smallskip}
LOB2       & 22.5 & 1.72 & 0 & 0.02 & 0.21 \\
\noalign{\smallskip}
MOB        & 23.5 & 1.72 & 0.62 & 0.02 & 0.2 \\
\hline\noalign{\smallskip}
\end{tabular}
\tablefoot{
\change{\pex\ fixed parameters: $E_c = 200$ keV, $\cos\theta = 0.45$, solar abundances.} \\
}
\end{table}

\begin{figure}
\centering
\includegraphics[width=\hsize]{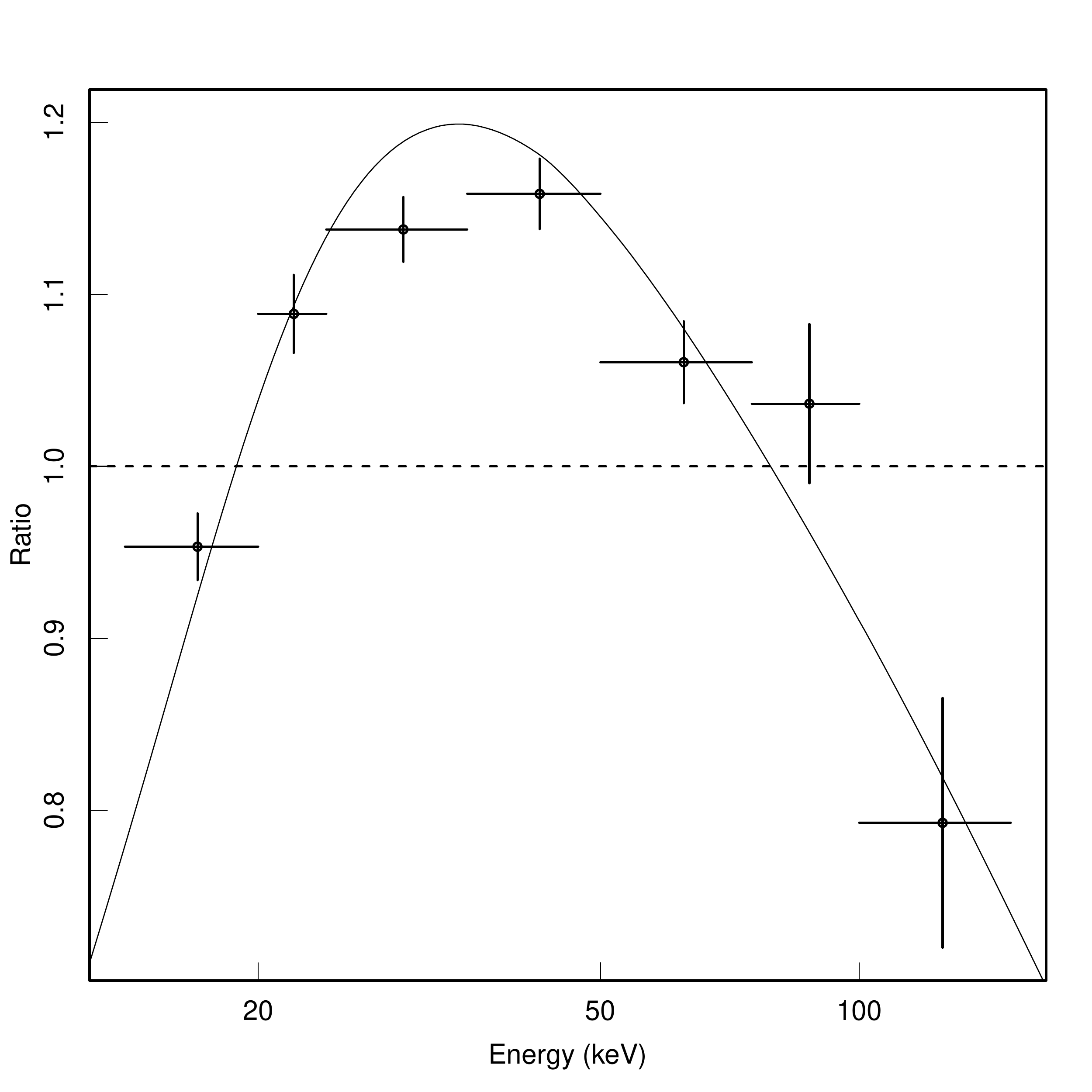}
\caption{\change{Ratio of MOB versus LOB1 spectra. The spectra are derived from the samples based on \Nh\ definition. The curve is the ratio between the two best fit models fitted with the \pex\ model.}}
\label{fig:ratiomoblob}
\end{figure}

We notice that the spectral index $\Gamma$ becomes softer for absorbed sources, in agreement with the results of \citet{2011ApJ...728...58B,2013ApJ...770L..37V}, who also analyzed Swift/BAT spectra stacked according to absorption. The mildly obscured samples (MOB Seyfert 2 and MOB) feature a strong curvature, signature of a high level of reflection (Tab. \ref{tab:spectralparameters}). \change{This reflection amplitude is larger than observed in the spectra derived for lightly absorbed samples and comparable with the reflection shown by the unabsorbed sample. The parameters listed in Table \ref{tab:spectralparameters} are model dependent and we do not know if e.g. the cutoff energy is the same in unabsorbed or absorbed sources, but the different curvatures are confirmed in Figure \ref{fig:ratiomoblob}, which is also not affected by uncertainties in the instrumental calibration and response, confirming that the difference in curvature observed between the LOB and MOB stacked spectra is not an artifact. This confirms with much higher signal to noise and solidity, previously obtained results \citep[R11,][]{2013ApJ...770L..37V}.}

This effect is not apparent in \citet{2011ApJ...728...58B} as they stacked only unabsorbed (\Nh\ $< 10^{22}~\text{cm}^{-2}$) and absorbed (\Nh\ $> 10^{22}~\text{cm}^{-2}$) spectra. We conclude hence that although the stacking method of \citet{2011ApJ...728...58B} and \citet{2013ApJ...770L..37V} are different from the one adopted in this work, the results are in agreement.

\change{The model dependence of the parameters in Table \ref{tab:spectralparameters} does not matter very much for the CXB synthesis as long as the spectral templates provide a good representation of the average spectra. We have verified that fitting the data with a different fixed cutoff energies provides different parameters but produce finally the same synthesied CXB. This will be discussed in more detail in Sec. \ref{sec:discussion}.}\\

To generate the spectral template of CTK sources we adopted the \myt\ model \citep{2012MNRAS.423.3360Y}, which properly calculate the absorption due to Compton scattering. This model assumes a powerlaw emission as primary component and a donut shaped torus with fixed solar abundancies surrounding the central emitter. It provides a transmitted component and a scattered component, the latter ones plays the role of the \pex\ reflection component. The shapes predicted by the \pex\ reflected component and the \myt\ scattered component are significantly different for \logNh $\geq 24~\text{cm}^{-2}$.

The transmitted and the scattered components of the \myt\ model can be separately fitted in order to measure the relative strength of the scattered component. As our template need to represent an average spectra, we fixed this relative strength to 1. We use as a primary component a cutoff powerlaw with $\Gamma = 1.9$ and $E_c = 200$ keV and the transmitted component is modeled with a powerlaw truncated at 200 keV with the same spectral index scattered by the absorbing torus. The inclination angle is fixed at $90^{\circ}$.

We introduced two templates for Compton Thick sources, separating them into moderately obscured ($10^{24} < \text{N}_\text{H} < 10^{25}~\text{cm}^{-2}$) and deeply obscured CTK (\Nh~$>~10^{25}~\text{cm}^{-2}$), as done in previous works \citep{2007A&A...463...79G,2014ApJ...786..104U}. The parameters used for the two templates are the same except for the absorption.\\

To conclude the template modeling, we finally added a \changeb{scattered component in the range $1 - 10$ keV} in obscured spectra (i.e. with \logNh\ $> 21$), following the recipe of \citet{2007A&A...463...79G}, to model the X-ray emission in excess of the absorbed powerlaw commonly observed in Seyfert 2 galaxies. It is modeled as a cutoff powerlaw with $E_c = 5$ keV, the same spectral index of the main component, and normalization fixed at 3\% of the normalization of the main component. As the iron line and other relevant emission lines are located at energies below the BAT lower energy threshold, we did not include any line in our modeling.

\begin{figure}
\centering
\includegraphics[width=\hsize]{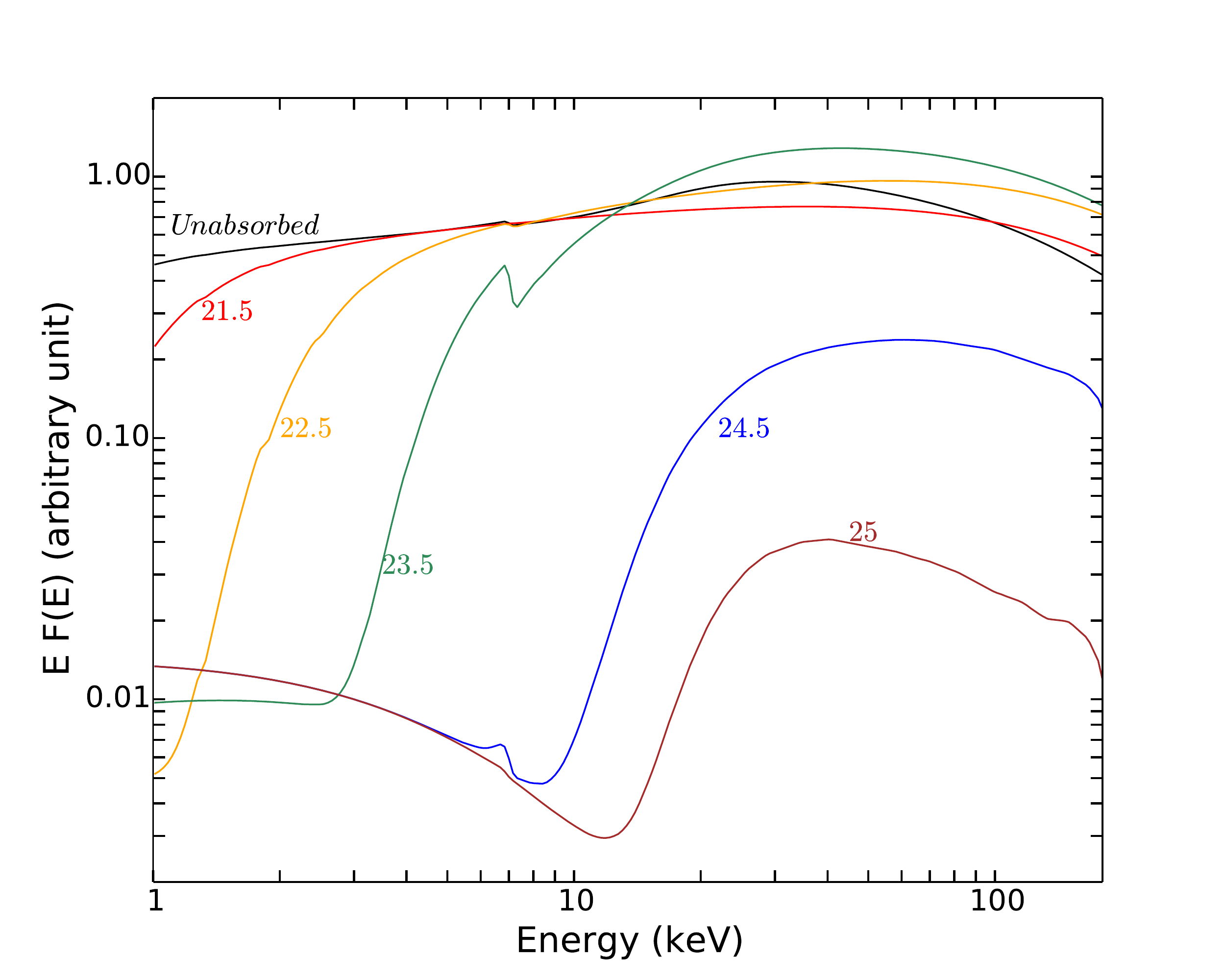}
\caption{\change{Spectral templates generated with the parameters presented in Table \ref{tab:spectralparameters}, stacked spectra based on \Nh. The templates are normalized such that the integrated emission of their primary component in the $2 - 10$ keV band is equal to 1. The numbers identify the templates by their \logNh\ value. The parameters of the CTK templates (\logNh $> 24$) are given in the text.}}
\label{fig:spectraltemplates}
\end{figure}

With the spectral parameters, we generated spectral templates in the range $1 \div 500$ keV (Fig. \ref{fig:spectraltemplates}) which were used for the CXB synthesis. Previous works, e.g. \citet{2007A&A...463...79G,2014ApJ...786..104U}, assumed a gaussian dispersion of spectral indexes. As our templates are derived from stacked spectra, they represent the average value of the sample, hence we choose to not include any dispersion.

We should point out that these spectral templates are derived by stacking local AGN. We assume that they are representative of AGN up to z $\sim 1$.

\section{X-ray luminosity function and \Nh\ distribution\label{sec:xlf}}

The XLF gives the AGN number density per comoving volume as a function of intrinsic unabsorbed luminosity and redshift. We used the XLFs of \citet{2003ApJ...598..886U}, \citet{2005A&A...441..417H}, \citet{2014ApJ...786..104U} and \citet{2015ApJ...804..104M} (U03, H05, U14 and M15 respectively). U03, U14 and M15 built their XLFs in the $2 - 10$ keV energy band, for both \mbox{``type-1''} and \mbox{``type-2''} objects, and using samples of 247, 4039 and $\sim 3200$ AGN, respectively. The XLF of H05 were built in the $0.5 - 2$ keV energy band and used 944 AGN \mbox{``type-1''} objects. \citet{2007A&A...463...79G} (G07) compared the H05 XLF with other XLFs \citep[U03,][]{2005ApJ...635..864L} built from \mbox{``type-1''} plus \mbox{``type-2''} objects in order to estimate a luminosity dependent ratio of obscured versus unobscured object $R_{o/u}(L_X)$, and corrected the H05 XLF in order to obtain the density of obscured AGN. We used the same correction.

The XLFs are derived fitting the sample with an analytical empirical \emph{smoothed two powerlaw formula} which has been found to fit well the data in the local universe. This expression is then corrected for a density evoluction factor $e_d$ to set the dependence on the redshift. The analytical expression of the density evolution factor is defined according to an evolution model. U03, H05 and U14 agreed that the best fit to the data is achieved using the Luminosity Dependent Density Evolution (LDDE) model, i.e. $e_d$ depends on both the X-ray luminosity $L_X$ and the redshift $z$, rather than other models like the Pure Luminosity Evolution (PLE) model where the density evolution factor depends on the redshift $z$ only. We used the best fit XLF models provided by them, which are described as:

\begin{equation}
\label{eq:xlfanalytic}
\frac{d \Phi \left (L_X, z \right )}{d \log L_X} = A \left [ \left ( \frac{L_X}{L_*} \right )^{\gamma_1} + \left ( \frac{L_X}{L_*} \right )^{\gamma_2} \right ]^{-1} e_d \left ( L_X, z \right )
\end{equation}

\noindent and the density evolution factor:

\begin{equation}
\label{eq:ed}
e_d \left ( L_X, z \right ) =
\begin{cases}
\left ( 1 + z \right )^{p_1} & \mbox{if} \; z \leq z_c \\
\left ( 1 + z_c \right )^{p_1} \left [ \left ( 1 + z \right ) / \left ( 1 + z_c \right ) \right]^{p_2} & \mbox{if} \; z > z_c
\end{cases}
\end{equation}

\noindent with 

\begin{equation}
\label{eq:zc}
z_c \left ( L_x \right ) =
\begin{cases}
z_c^{*} \left (L_X / L_{X,c} \right )^\alpha & \mbox{if} \; L_X \leq L_{X,c} \\
z_c^{*} & \mbox{if} \; L_X > L_{X,c}
\end{cases}
\end{equation}

While U03 used exactly these equations, H05 introduced an extra smooth dependence on $L_X$ in Eq. \ref{eq:ed}, in the indexes $p_1$ and $p_2$, while U14 introduced an extra variation in the slope at high redshift in Eq. \ref{eq:ed}, having hence three slopes, $p_1$, $p_2$ and $p_3$, and two redshift thresholds, $z_{c,1}^{*}$ and $z_{c,2}^{*}$. M15 used the same definition of U14, but fixed different parameter in the fit. The differences between these XLFs are due not only to the dataset used, but also to these small but likely relevant different parametrizations of the analytical equations.

In the CXB synthesis process, the analytic equation of the XLF is multiplied by the luminosity $L_X$ and then integrated in the range $42 < \log L_x < 48$ and $0 < z < 5$. \change{The uncertainties of the XLFs at high luminosity and high redshift are not relevant for the CXB synthesis as the contribution to the CXB of AGN in this range is negligible.}\\

\begin{figure}
\centering
\includegraphics[width=\hsize]{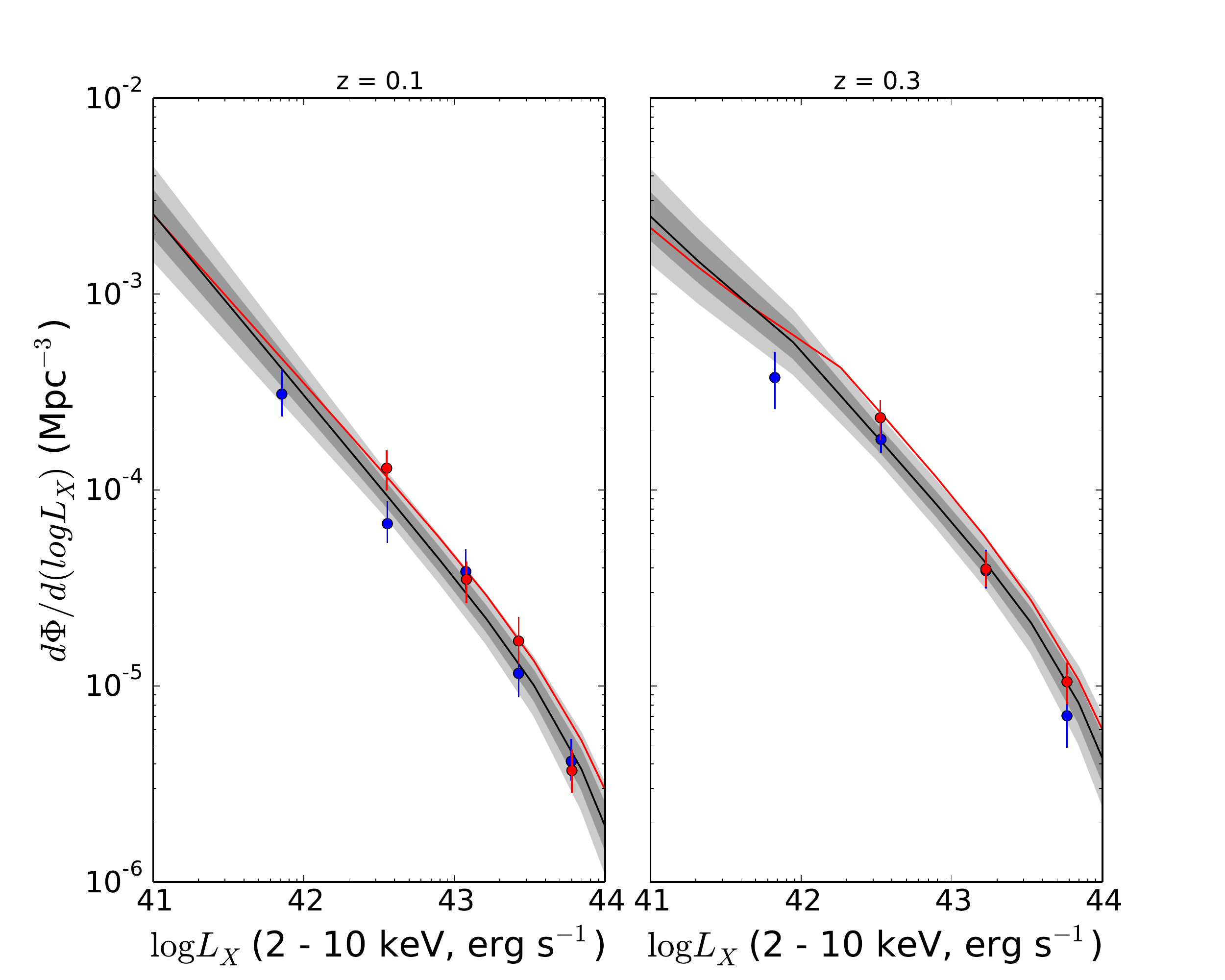}
\caption{XLF of U14 plotted at redshift z = 0.1 and z = 0.3 in the range $41 < \log L_x < 44$. The black line represent the best fit, gray regions show the effect of uncertainties in the parameters: dark (light) gray correspond to 1$\sigma$ (2$\sigma$) region. The red line is U03 XLF. Red and blue points are taken from U14, and represent the data derived from the soft (< 2 keV) and hard (> 2 keV) X-ray band respectively, \change{converted in 2 to 10 keV luminosities as explained in U14.}}
\label{fig:xlfu14}
\end{figure}

Figure \ref{fig:xlfu14} shows the U14 XLF for two redshift bins with the confidence regions at 1$\sigma$ and 2$\sigma$ and the U03 XLF for comparison. The confidence region is mostly related to the uncertainties on $\gamma_1$ and $L_*$. The uncertainty on the XLF normalization can be up to $\sim 10\%$ (1$\sigma$). The CXB synthesis of U14 (whose XLF is mostly based on Chandra surveys) falls 10 - 20\% short of the CXB flux measured by Chandra. This indicates that the XLF is probably underestimated even though other sources can contribute to the CXB in the soft X-ray band.\\

The fraction of AGN with different \Nh\ must be known to synthetize the CXB. We used the \Nh\ distribution derived by G07 and \citet{2009ApJ...696..110T} (T09) and also derived it to obtain the best match between the observed and syntetized CXB. These \Nh\ distributions, displayed in Figure \ref{fig:nhdistr} show significant discrepancies.

It has to be pointed out that these \Nh\ distributions do not depend on redshift and luminosity, contrasting with some observations \citep[U14]{2005ApJ...635..864L,2008A&A...490..905H,2010MNRAS.401.2531A,2015MNRAS.451.1892A,2015ApJ...804..104M,2015ApJ...802...89B}. To account for this effect, U14 proposed an empirical analytical function for the \Nh\ distribution dependent on $L_X$ and $z$. We tested that function togheter with the U14 and M15 XLFs, which used it in their works.

The CTK AGN are considered separately from the above \Nh\ distributions, their fraction is evaluated adjusting the CXB synthesis to the observations.

\section{Synthesis of the CXB\label{sec:results}}

\subsection{Comparison with previous works\label{sec:synt_previous}}

\begin{figure*}
\centering
\includegraphics[width=\hsize]{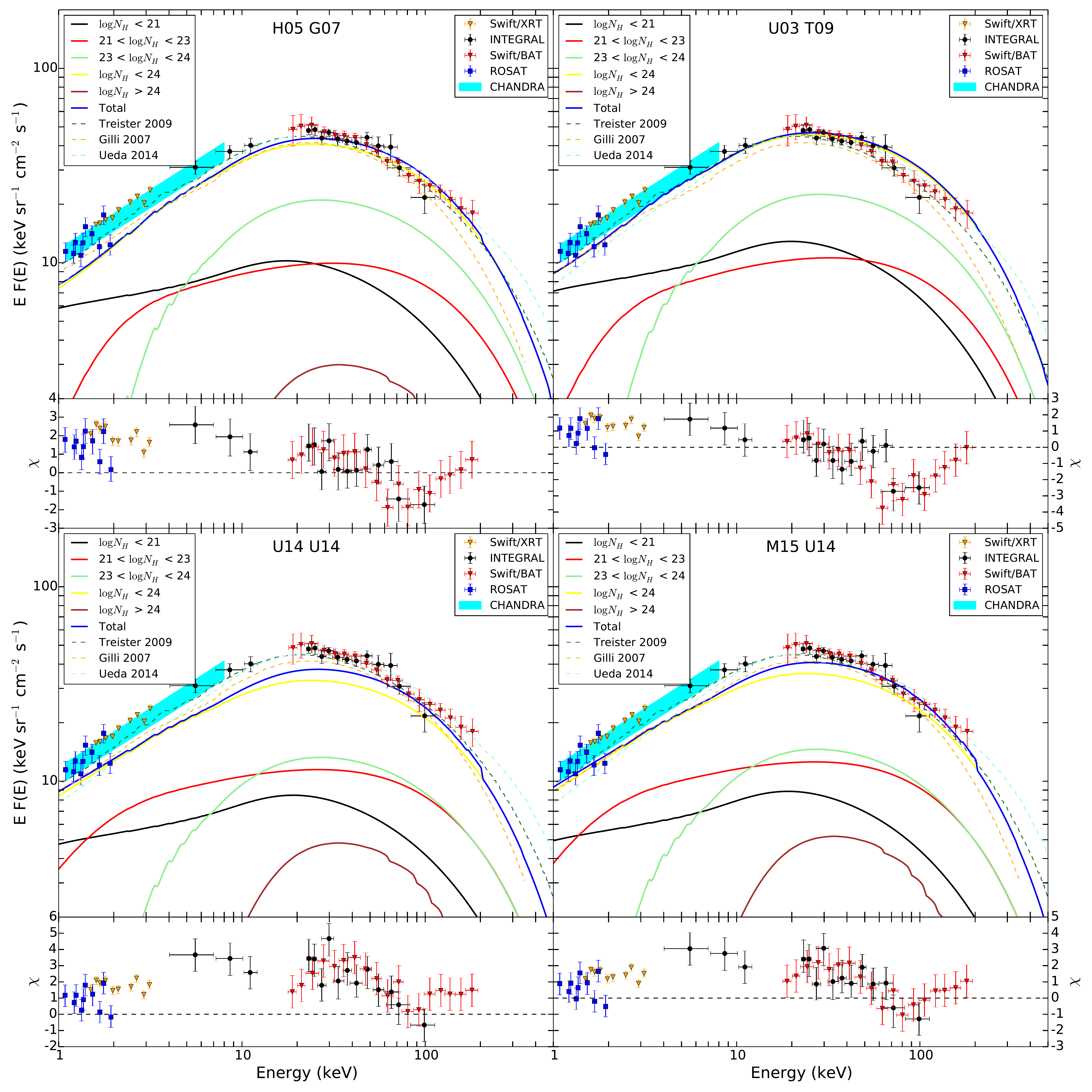}
\caption{Synthesis of the CXB (thick blue line) based on BAT spectral templates for different XLFs and \Nh\ distributions. The title of each panel quotes the papers from wich the XLFs and \Nh\ distributions respectively are taken from. The data from Swift/XRT (orange triangles), ASCA SIS (green squares) INTEGRAL (black points), Swift/BAT (red squares), HEAO1 (gray triangles), ROSAT (blue triangles) and Chandra (cyan region) are shown. Each panel shows the spectra and the difference between the data and the synthesis.}
\label{fig:multicxb}
\end{figure*}

\begin{table*}
\caption{Comparison of Compton thick estimates with previous works using BAT spectral templates derived from the samples based on \Nh. \change{For each paper we report both the CTK flux estimated in the original paper and the one estimated with the BAT templates, togheter with the $\chi^2_ {red}$ of the fit.}}
\label{tab:comparectk}
\begin{center}
\begin{tabular}{lccccccc}
\hline\noalign{\smallskip}
\hline\noalign{\smallskip}
\smallskip
Paper & XLF\tablefootmark{a} & \Nh\tablefootmark{b} & Templates & CTKR\tablefootmark{c} & Datasets\tablefootmark{d} & Flux\tablefootmark{e} & $\chi^2_{red}$\\
\hline\noalign{\smallskip}
G07 & H05 & G07 & G07 & $24 \div 26$ & ASCA SIS, HEAO1 & 10 & \\
 & & & BAT & & & $<0.3$ & > 2 \\
\hline\noalign{\smallskip}
T09 & U03 & T09 & T09 & $24 \div 25$ & XMM, CHANDRA, BAT, INTEGRAL & 4 & \\
 & & & BAT & & & $<1$ & > 2 \\
\hline\noalign{\smallskip}
U14 & U14 & U14 & U14 & $24 \div 26$ & ASCA SIS, BAT, INTEGRAL & $9 \pm 4$ & \\
 & & & BAT & & & $9.5 \pm 0.5$ & 1.74 \\
\hline\noalign{\smallskip}
\end{tabular}
\end{center}
\tablefoot{
\tablefoottext{a}{XLF used in the corresponding paper.} \\
\tablefoottext{b}{\Nh\ distribution used in the corresponding paper.} \\
\tablefoottext{c}{\logNh\ range of CTK AGN considered in the synthesis.} \\
\tablefoottext{d}{Datasets used to compare with the CXB synthesis.} \\
\tablefoottext{e}{Flux at the peak of the CTK component in the synthesis, in unit of keV cm$^{-2}$ s$^{-1}$ Str$^{-1}$.} \\
}
\end{table*}

Figure \ref{fig:multicxb} shows different synthetized spectra obtained selecting combinations of XLF and \Nh\ distribution used in previous works (G07, T09, U14 and M15) and using the set of templates derived from Swift/BAT. These spectra are plotted against Swift/XRT \citep{2009A&A...493..501M}, ASCA SIS \citep{1995PASJ...47L...5G}, INTEGRAL \citep{2007A&A...467..529C}, Swift/BAT \citep{2008ApJ...689..666A}, HEAO1 \citep{1999ApJ...520..124G}, ROSAT \citep{1998A&A...334L..13M} and \mbox{Chandra} \citep{2006ApJ...645...95H} data. It should be pointed out that we do not perform any fit or renormalization against the data. Figure \ref{fig:multicxb} shows also the separated contribution of unabsorbed, LOB (here is the sum of \logNh\ = 21.5 and \logNh\ = 22.5), and MOB AGN. The CTK fraction is usually estimated to fill the gap between the total contribution of Compton thin objects and the data. The CTK fraction shown corresponds to the one estimated in the corresponding paper.

The main differences between the synthesis models are caused by the XLF. Even without CTK sources, H05 and U03 XLFs are in good agreement with the data abobe 30 keV. The U14 XLF instead produces a CXB spectrum, 10 - 20\% fainter than that observed by Chandra, with a wide gap between the data and the spectrum at $\sim 30$ keV. As mentioned previously, this XLF is too faint to account for the CXB at soft X-rays, and requires a large fraction of CTK sources to match the CXB at hard X-rays.

\subsection{Fitting the Compton Thick contribution\label{sec:ctkcontribution}}

The CTK contribution to the CXB is estimated by varying the CTK fraction through a $\chi^2$ minimization. To investigate the effect of using the BAT templates (Sec. \ref{sec:templates}), the CXB is synthetized with the combinations of XLFs and \Nh\ distributions used by G07, T09 and U14 in order to estimate the CTK fraction. We used the same CXB datasets used in the respective papers. Table \ref{tab:comparectk} reports the flux due to the CTK sources at the peak of the integrated spectrum. With the BAT templates, we obtain significantly less CTK sources. However in all these cases the fit to the CXB is not acceptable. Table \ref{tab:comparectk} refers to results obtained with the BAT spectral templates derived from the samples based on absorption, using the samples based on Seyfert type leads to similar results.

We then compared the CXB synthesis with the CXB observed with ROSAT, Swift/XRT, Swift/BAT and INTEGRAL. \change{We considered only the ROSAT data above 1 keV to exclude the soft excess component from the CXB modeling.} These datasets as well as those of XMM-Newton \citep{2009A&A...493..501M} and Chandra are well cross-calibrated and feature similar CXB normalization. Their calibration also matches these of the recent XLF obtained with deep soft X-ray surveys by XMM and Chandra.

\begin{table}
\caption{Compton thick estimates obtained using BAT spectral templates derived from the samples based on absorption, fitting only the CTK density, for various combination of XLFs and \Nh\ distributions. The datasets used in the fitting process are ROSAT, Swift/XRT, Swift/BAT and INTEGRAL.}
\label{tab:ctkfit}
\begin{center}
\begin{tabular}{ccccc}
\hline\noalign{\smallskip}
\hline\noalign{\smallskip}
\smallskip
XLF & \Nh & Flux\tablefootmark{a} & $f_{CTK}$\tablefootmark{b} & $\chi^2_{red}$ \\
\hline\noalign{\smallskip}
H05 & G07 & $3.0 \pm 0.5$ & $24\%$ & 2.12 \\
\noalign{\smallskip}
U03 & T09 & $0^{+1.0}_{-0}$ & $< 5\%$ & 1.92 \\
\noalign{\smallskip}
U14 & U14 & $11.5 \pm 0.5$ & $47\%$ & 1.92 \\
\noalign{\smallskip}
M15 & U14 & $8.0 \pm 0.5$ & $47\%$ & 1.64 \\
\noalign{\smallskip}
\hline\noalign{\smallskip}
\end{tabular}
\end{center}
\tablefoot{
\tablefoottext{a}{Flux at the peak of the CTK component in the synthesis, in unit of keV cm$^{-2}$ s$^{-1}$ Str$^{-1}$.} \\
\tablefoottext{b}{Fraction of CTK (\logNh\ $> 24$) over all AGN (Any \Nh, CTK themselves included).}
}
\end{table}

Table \ref{tab:ctkfit} shows the estimates of the CTK fraction for the combinations of XLFs and \Nh\ distributions shown in Figure \ref{fig:multicxb}. We report both the flux of the CTK contribution at the peak and the fraction of CTK objects (assuming that all CTK are in the \Nh\ range $10^{24} - 10^{25}~\text{cm}^{-2}$) needed to produce this flux, as well the best $\chi^2_{red}$. With the XLF and \Nh\ distribution used by G07 and T09 and with the BAT templates the amount of CTK allowed by the synthesis is less than in the referred papers confirming that BAT templates produce higher CXB flux. Even with the XLF and \Nh\ distribution of U14 we obtain a lower CTK fraction. The fit to the CXB data is not good whatever are the combitation of XLF and \Nh\ distribution considered: the BAT templates with the previous models can not adequately represent the data.

\subsection{Fitting the \Nh\ distribution\label{sec:fittingnh}}

In order to achieve a better representation of the CXB spectrum, we adjusted the \Nh\ distribution directly to the data. We performed several fits with different assumptions to investigate possible systematics effects: we used the XLF of U14 or U03, we fixed the maximum \Nh\ of the distribution at $10^{25} ~\text{cm}^{-2}$, as done by T09, or at $10^{26} ~\text{cm}^{-2}$, as done by G07 and U14 (the implications of this choice are discussed later), and we used the BAT templates based on the samples defined according to \Nh\ or Seyfert type (see Sec. \ref{sec:stacking}).

There are several sources of uncertainties on the CXB normalization and synthesis. We improved the uncertainty on the spectral templates by using average hard X-ray spectra. The integrated XLF varies by 15\% for different assumptions and by 13\% taking $1 \sigma$ statistical uncertainties on the XLF into account. The CXB normalization between the various datasets varies by 10 - 15\% and the flux of the CXB synthesis produced with the U14 XLF falls 10 - 20\% below the Chandra CXB data. \change{In addition, we investigated the fraction of absorbed AGN which still escape detection. We simulated spectra with \Nh\ uniformly distributed into 6 linearly spaced bins in the range $10^{23} - 10^{24}~\text{cm}^{-2}$ (corresponding to our MOB sources definition), with the spectral index $\Gamma = 1.72$ derived for the MOB sources (Tab. \ref{tab:spectralparameters}), and convolving them with the Chandra matrix response assuming the total exposure of the Deep Chandra Surveys used by U14. We then looked for the deabsorbed luminosity required to be detected by Chandra, and integrated U14 XLF from to the lower limit ($\log L_x = 42$) to this luminosity threshold in order to estimate the AGN fraction which fails to be detected. We estimated that 11\% of the mildly obscured AGN at redshift < 1.3 have too faint spectra to be detected even in deep Chandra observations, adding an additional source of uncertainty to the XLF normalization.}

In order to account for these uncertainties in the fitting process we introduced a scaling factor wich allow for a renormalization of the synthesis on the data. The resulting factors vary from 1 to 1.3 for different XLFs.

\begin{figure}
\centering
\includegraphics[width=\hsize]{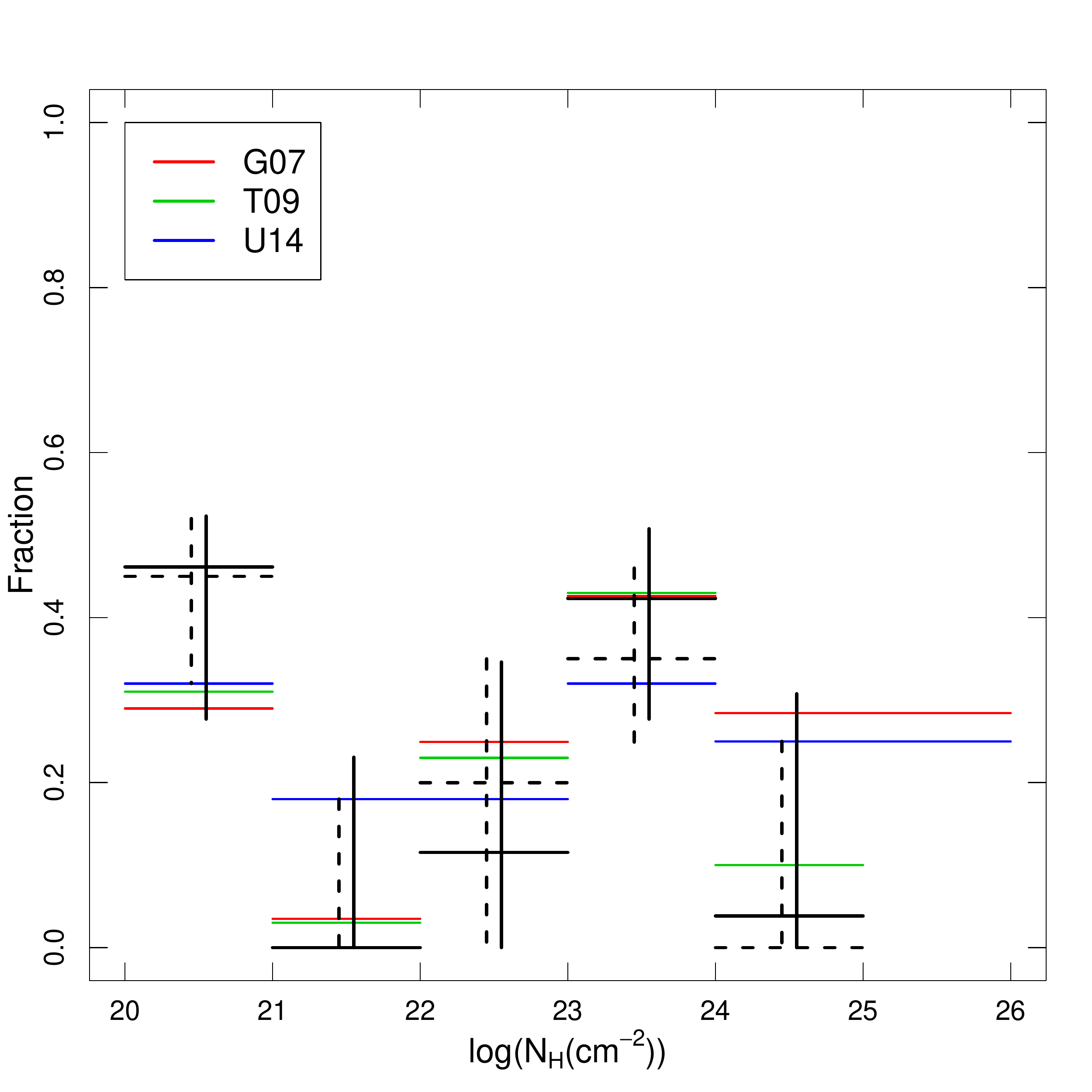}
\caption{Best fit \Nh\ distribution with U14 XLF (continuous black lines) and U03 (dashed black lines) compared with \Nh\ distribution from previous works. Error bars show the 1 $\sigma$ confidence region. To better compare with previous works, here the \Nh\ distributions are renormalized up to \Nh~$= 10^{24}~\text{cm}^{-2}$.}
\label{fig:nhdistr}
\end{figure}

Figure \ref{fig:nhdistr} shows the best fit \Nh\ distributions using the XLF of U14 (continuous black lines) and U03 (dashed black lines), compared with the \Nh\ distributions of G07, T09 and U14. The uncertainties in the fit of the \Nh\ distributions are such that they are compatible with the \Nh\ distributions published previously.

\begin{table}
\caption{Compton thick estimates obtained fitting the whole \Nh\ distribution. The datasets used for the fit are ROSAT, Swift/XRT, Swift/BAT and INTEGRAL. \change{We tested different combination of XLFs, \logNh\ range of the CTK AGN, and templates (based on spectra stacked according to \Nh\ or R11 definition. We report the scaling factor found, Flux of the CTK component, CTK fraction and $\chi^2_{red}$.}}
\label{tab:nhdistrfit}
\begin{center}
\begin{tabular}{ccccccc}
\hline\noalign{\smallskip}
\hline\noalign{\smallskip}
\smallskip
XLF & CTKR & Templates & SF\tablefootmark{a} & Flux\tablefootmark{b} & $f_{CTK}$ & $\chi^2_{red}$ \\
\hline\noalign{\smallskip}
U14 & $24 \div 25$ & BAT \Nh & 1.3 & $0.4^{+2.9}_{-0}$ & $<25\%$ & 0.99 \\
\noalign{\smallskip}
U14 & $24 \div 25$ & BAT R11 & 1.3 & $0^{+3}_{-0}$ & $<22\%$ & 1.10 \\
\noalign{\smallskip}
U14 & $24 \div 26$ & BAT \Nh & 1.3 & $0.5^{+2.9}_{-0}$ & $<36\%$ & 0.99 \\
\noalign{\smallskip}
U03 & $24 \div 25$ & BAT \Nh & 1 & $0^{+2.1}_{-0}$ & $<12\%$ & 1.17 \\
\noalign{\smallskip}
\hline\noalign{\smallskip}
\end{tabular}
\end{center}
\tablefoot{
\tablefoottext{a}{Scaling Factor.}\\
\tablefoottext{b}{Flux at the peak of the CTK component in the synthesis, in unit of keV cm$^{-2}$ s$^{-1}$ Str$^{-1}$.}
}
\end{table}

\begin{figure}
\centering
\includegraphics[width=\hsize]{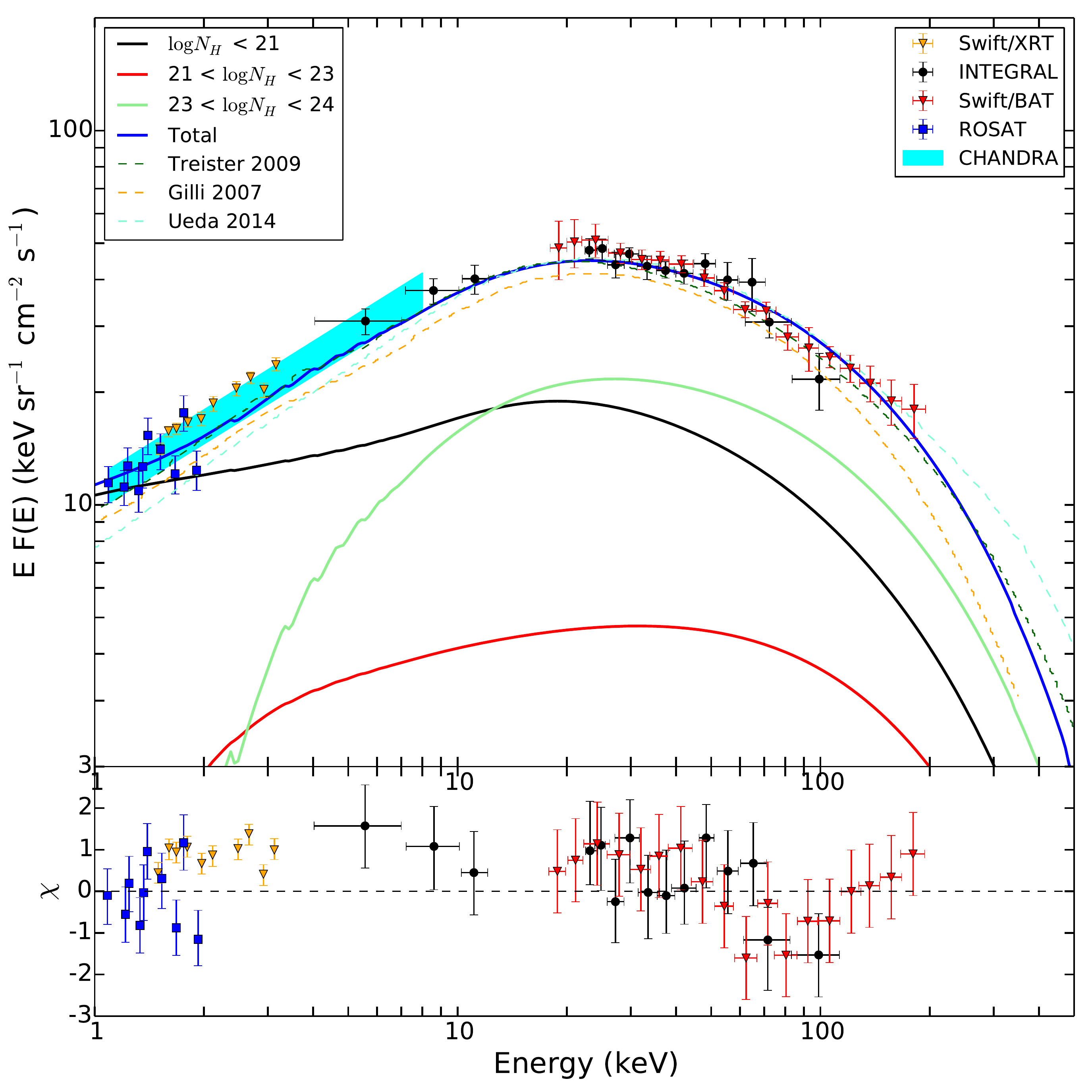}
\includegraphics[width=\hsize]{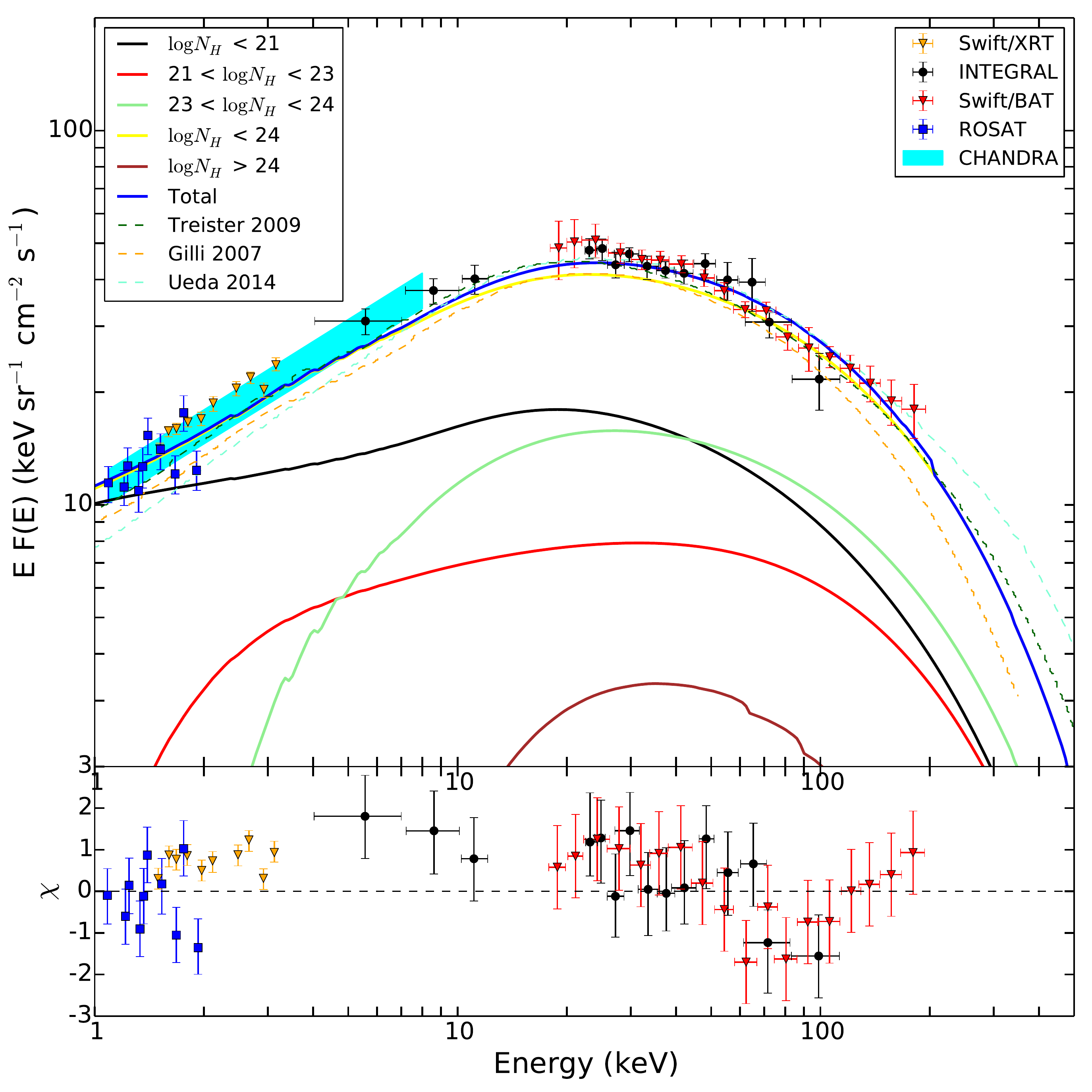}
\caption{\emph{Top panel:} Synthesis of the CXB spectrum and residuals (in unit of $\chi^2$) obtained using the best fit \Nh\ distribution derived with the selection of Table \ref{tab:nhdistrfit}, row 1. \emph{Bottom panel:} Same a the top banel, but using the \Nh\ distribution corresponding to the 1 $\sigma$ upper limit of the CTK fraction. Previously published CXB synthesis from T09, G07, U14 (dotted lines) are also shown.}
\label{fig:cxbbestfit}
\end{figure}

Table \ref{tab:nhdistrfit} shows the CTK fraction estimated for several XLFs and CTK bin sizes. Figure \ref{fig:cxbbestfit} shows the CXB synthesis obtained with the \Nh\ distribution (best fit and 1 $\sigma$ upper limit for the CTK fraction) corresponding to the first row of Table \ref{tab:nhdistrfit}. All the above combinations are able to represent the data providing acceptable $\chi^2_{red}$. Table \ref{tab:nhdistrfit} shows that the scaling factor is always 1.3 with the U14 and 1 wth the U03 XLF. We consider that such scaling factors are still acceptable considering the uncertainties on the data, on the XLF parameters and on the unresolved sources.

\section{Discussion\label{sec:discussion}}

\subsection{Stacked BAT spectra\label{sec:spectra}}
\begin{figure}
\centering
\includegraphics[width=\hsize]{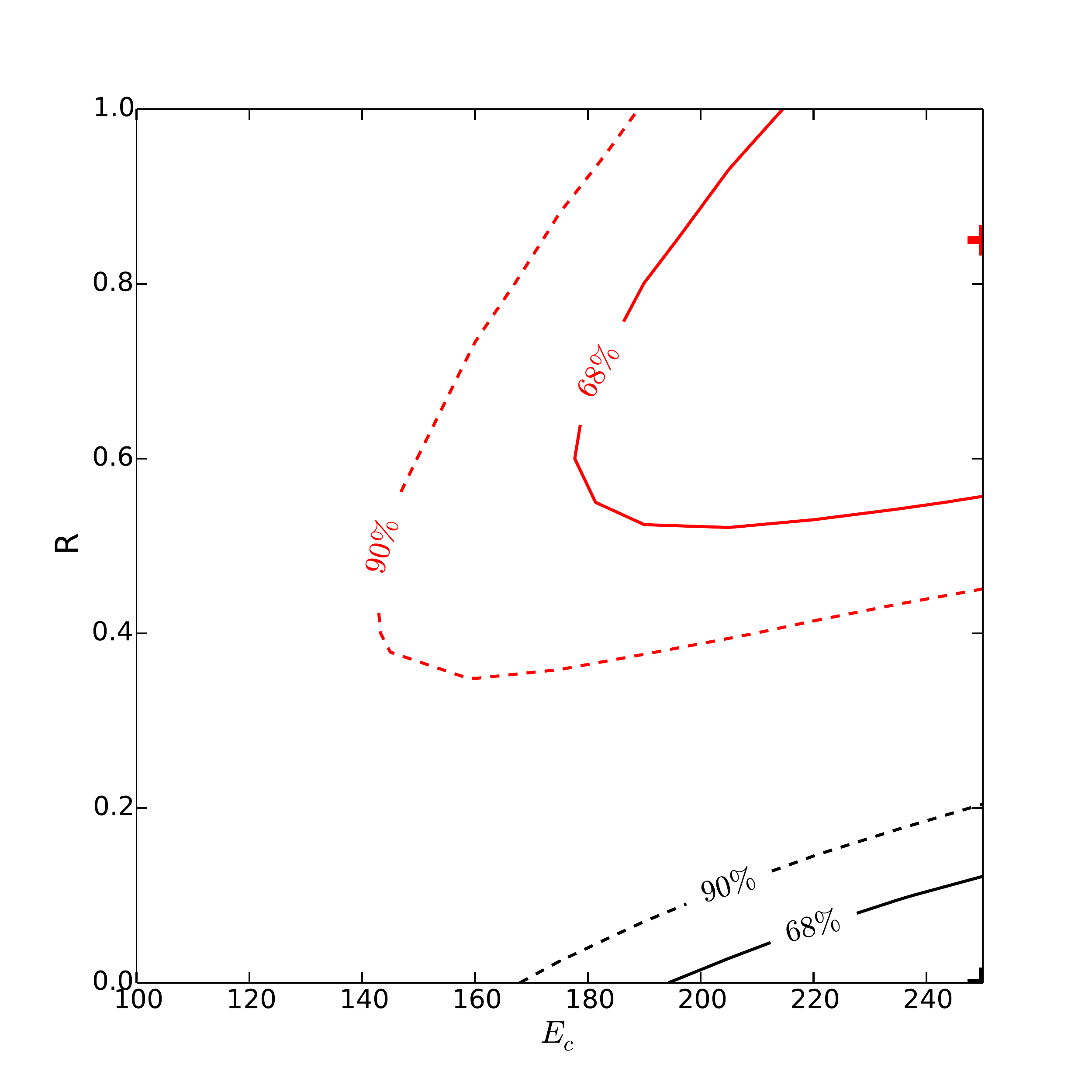}
\caption{\change{Contour plot of the $\chi^2$ as a function of $E_c$ and R for the LOB1 ($10^{21} < \text{N}_\text{H} < 10^{22}~\text{cm}^{-2}$) spectrum (black) and the MOB $10^{23} < \text{N}_\text{H} < 10^{24}~\text{cm}^{-2}$) spectrum (red). Continuous lines show the 1$\sigma$ (68\%) confidence region, dotted lines show the 90\% confidence region. The crosses mark the best fit position.}}
\label{fig:EcRcontour}
\end{figure}

The unified model assumes that different classes of AGN can be explained by the observer line of sight crossing or not an anisotropic absorbing medium \citep{1993ARA&A..31..473A,1995PASP..107..803U}. This does not explain why MOB sources feature a larger reflection. A clumpy absorber model can offer a solution, as obscuration and reflection could increase with the number of clumps. This geometry was put forward independently to explain high resolution infrared spectra of AGN \citep{2011ApJ...731...92R,2012ApJ...747L..33E}.

The BAT spectra of Compton thin AGN are modeled with the \pex\ model with a cutoff energy fixed at 200 keV. There is a degeneracy between the cutoff energy and the reflection. Figure \ref{fig:EcRcontour} shows the confidence contours as a function of $E_c$ and R for the spectra of the LOB1 and MOB samples respectively, and indicates that the stronger reflection observed in MOB sources is solid and not an effect of the fixed cutoff energy.

To verify the effect of the fixed cutoff energy on the CXB synthesis, we also derived the best fit parameters for $E_c = 150$ keV and $E_c = 250$ keV and performed the synthesis with these templates. The bump at $\sim 30$ keV is well reproduced in all cases and the templates contribute to the synthesis in the same way, because the spectral models remain a good representation of the data. Changing the cutoff energy does not significantly change the spectral index, therefore the flux of the template in the energy band of the XLF ($2 - 10$ keV for U14) is not significantly different for these cases, leading to similar normalizations in the CXB synthesis.

As mentioned in Section \ref{sec:templates}, previous works (e.g. G07, U14) introduced a gaussian dispersion on the spectral index of their templates \changeb{to reproduce the variance of the observed spectral slopes}. The effect on the synthesis is to increase the flux of the CXB at hard X-rays (G07). We checked the effect of adding a spectral index dispersion in our synthesis and noticed that in our case the effect is not as strong \changeb{as the BAT spectra already measure the spectral shape in the range 15 to 150 keV.} We therefore did not introduce a spectral index distribution also as the BAT templates \changeb{already average the contribution of many sources.}

\subsection{Contribution of Compton Thick AGN\label{sec:ctkcontribute}}

\begin{figure}
\centering
\includegraphics[width=\hsize]{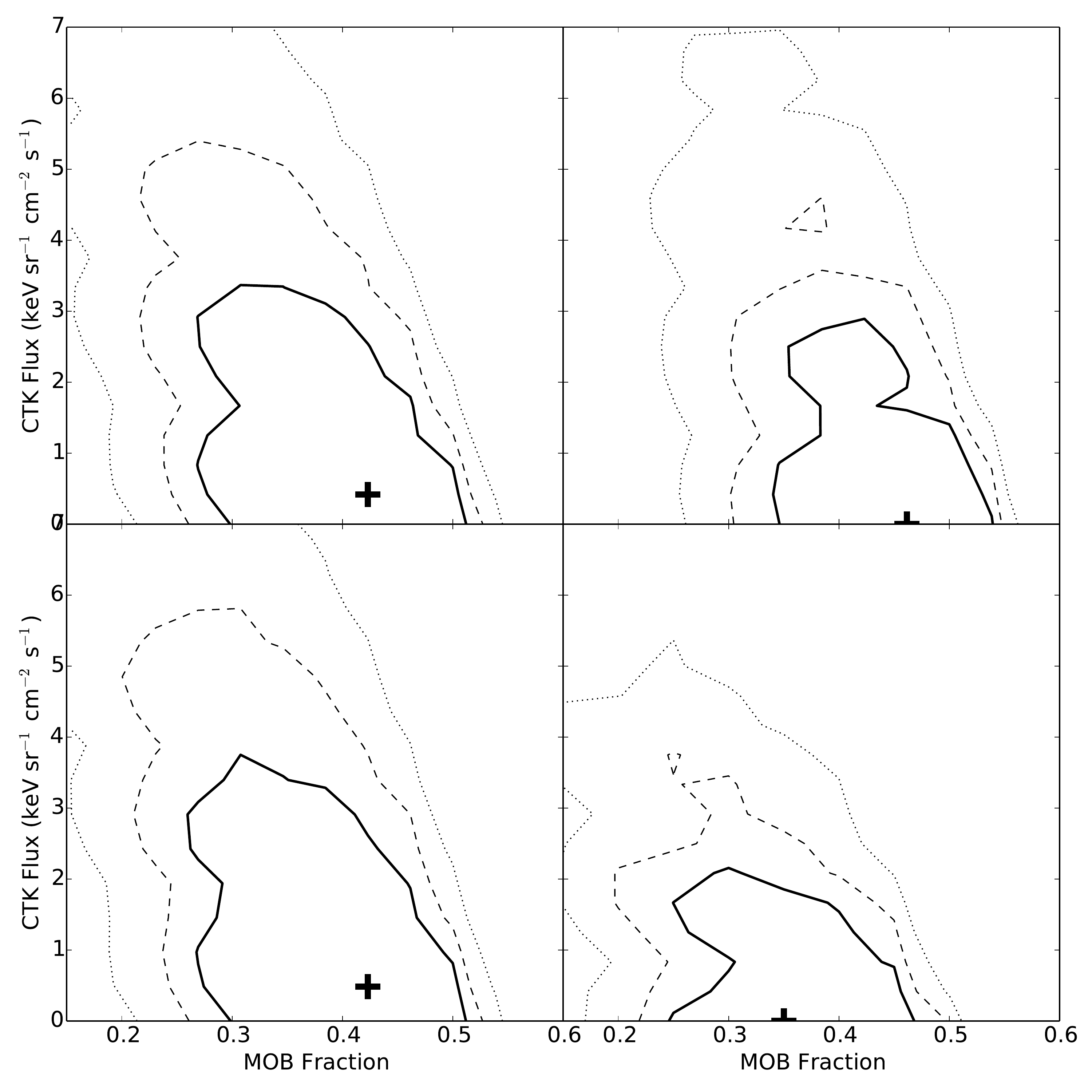}
\caption{\change{Contour plot of MOB fraction versus the Flux of the CTK contribution in the CXB at 30 keV} (peak of the integrated spectrum) for the different assumptions used in Table \ref{tab:nhdistrfit}: \emph{Top left:} row 1, \emph{Top right:} row 2, \emph{Bottom left:} row 3, \emph{Bottom right:} row 4. The crosses mark the best fit position. Continuous, dashed and dotted lines represent the 68\% (1 $\sigma$), 90\% and 99\% confidence region respectively.}
\label{fig:contourmobctk}
\end{figure}

Figure \ref{fig:contourmobctk} shows the contour plot of the fraction of MOB sources and of the flux at the peak of the CTK contribution for the various models listed in Table \ref{tab:nhdistrfit}. The scaling factor added in the fitting procedure is taken into account. All the contour plots show the expected tendency that the fraction of MOB decreases in order to allow for a stronger CTK contribution. For all the combinations investigated, the best fit indicates a CTK flux at the peak less than 1 keV cm$^{-2}$ s$^{-1}$ Str$^{-1}$ and a 1 $\sigma$ upper limit of 2.1 - 2.9 keV cm$^{-2}$ s$^{-1}$ Str$^{-1}$ for the U03 and respectively U14 XLF, corresponding to 4 - 6\% of the CXB emission. These fluxes correspond to a CTK emission two (T09) to three times (U14 and G07) fainter than that estimated in previous works.

Our fluxes correspond to a maximum fraction of CTK AGN between 12\% (for U03 XLF) and 21\% (for U14 XLF) assuming that all CTK objects have \Nh\ $< 10^{25}~\text{cm}^{-2}$. As CTK objects are mostly undetected, it is not clear where the \Nh\ distribution should end. In the CXB modeling, the maximum absorption of CTK sources is typically considered to be \Nh~$=10^{25}~\text{cm}^{-2}$ (e.g. T09) or \Nh~$=10^{26}~\text{cm}^{-2}$ (e.g. G07, U14). The CTK fraction thus depends on the \Nh\ distribution of the CTK sources, which is unknown: previous works simply assumed it to be constant. \change{Adding a separate spectral template for deeply obscured CTK sources (25~<~\logNh~<~26) in our fitting procedure, we found that their fraction is basically unconstrained by the data. Fixing it to the same value as of mildly obscured CTK sources we get a CTK fraction between 21\% (for U03 XLF) and 29\% (for U14 XLF).}

\changeb{In the fitting procedure of the \Nh\ distribution (Sec. \ref{sec:fittingnh}) we added a scaling factor in order to account for various systematics effect. We did not introduce this scaling factor when fitting the Compton Thick contribution alone (Sec. \ref{sec:ctkcontribution}) because there we wanted to use exactly the same hypotheses as used in previous works, in order to verify the effect of the BAT templates on the CTK fraction estimation. We found that using these assumptions and the measured hard X-ray templates, does not provide a good fit to the CXB.}

\change{In the Swift/BAT catalogue of local AGN, only 5.5\% of them are classified as CTK (\logNh~>~24) \citep{2011ApJ...728...58B}}, while in the CHANDRA South Deep Field this fraction goes up to 20\% \citep{2012MNRAS.423..702B}. The discrepancy is related to the different sensitivity of the two instruments: \citet{2011ApJ...728...58B} estimated that correcting for the bias against detection of very absorbed sources the intrinsic CTK AGN fraction over the total AGN is 20\%, in agreement with \citet{2012MNRAS.423..702B}. As most of these CTK objects have \Nh\ $< 10^{25}~\text{cm}^{-2}$ this is consistent with our upper limit of 21\%. We hence conclude that a population of Compton thick AGN larger than that effectively observed is not required to account for the CXB.

The CXB spectrum is dominated by low luminosity sources, but also by low redshift sources: 99\% of the CXB flux is generated by AGN located at z < 1.3. The CXB hence can not be used to constrain galaxy evolution at high redshift. It is yet unclear if mergers of galaxies in the early Universe have triggered rapid black-hole growth, stellar formation and obscuration \citep{2006ApJS..163....1H} or if the evolution has been more linear \citep{2015ApJ...811..148C}. The stronger reflection observed in mildly obscured sources locally should be probed at higher redshift to study how the quasar environment evolved \citep{2013ApJ...773..125A}.

\section{Summary and Conclusions\label{sec:conclusions}}

We measured the averaged hard X-ray spectral properties of seveal samples of Seyfert galaxies by stacking BAT spectra. We found that mildly obscured sources ($10^{23}~<~\text{N}_\text{H} ~<~10^{24}~\text{cm}^{-2}$) feature a stronger reflection than less absorbed sources, suggesting that AGN are surrounded by a clumpy rather than by a donut shaped torus.

The stacked BAT spectra have been used to define spectral templates of Seyfert galaxies and to synthetize the diffuse cosmic X-ray background. We found that the strong reflection of mildly obscured sources contribute massively to the bulk of the CXB emission, leaving little space for the contribution of Compton thick sources, in contrast with the results of previous works. The fraction of Compton thick sources estimated from our synthesis is less than 21\%, compatible with that obtained from deep surveys.

We investigated possible systematic effects in the synthesis process due to assumptions in the modeling, using different XLF, spectral templates built on different samples, considering or not the contribution of strongly obscured Compton Thick sources ($10^{25}~<~\text{N}_\text{H} ~<~10^{26}~\text{cm}^{-2}$). In all cases, only 4 - 6\% of the flux of the CXB at 30 keV can be attributed to Compton Thick sources.

We allowed a renormalization factor on the absolute flux of the CXB synthesis to account for the uncertainties in the CXB measurements and in the XLF parameters. The MVN \change{(Monitor Vsego Neba)} X-ray astronomical experiment \citep{2012SPIE.8443E..10R} will help improving our knowledge on the CXB normalization.

\bibliographystyle{aa} 
\bibliography{bat_cxb} 

\end{document}